\documentclass[aps,twocolumn,showpacs]{revtex4}
\usepackage[dvips]{graphics}
\usepackage{setspace}
\usepackage{color}
\definecolor{gold}{rgb}{0.85,0.66,0}
\definecolor{dblue}{rgb}{0,0,0.8}

\usepackage[dvips]{graphics}

\begin{document}

\title{{\textcolor{gold}{Magnetic Quantum Wire as a Spin Filter: An 
Exact Study}}}

\author{{\textcolor{dblue}{Moumita Dey}}$^1$, {\textcolor{dblue}{Santanu 
K. Maiti}}$^{1,2}$ and {\textcolor{dblue}{S. N. Karmakar$^1$}}}

\affiliation{$^1$Theoretical Condensed Matter Physics Division,
Saha Institute of Nuclear Physics, 1/AF, Bidhannagar, Kolkata-700 064,
India \\
$^2$Department of Physics, Narasinha Dutt College, 129 Belilious Road,
Howrah-711 101, India}

\begin{abstract}
We propose that a magnetic quantum wire composed of magnetic and 
non-magnetic atomic sites can be used as a spin filter for a wide 
range of applied bias voltage. We adopt a simple tight-binding 
Hamiltonian to describe the model where the quantum wire is attached 
to two semi-infinite one-dimensional non-magnetic electrodes. Based 
on single particle Green's function formalism all the calculations 
are performed numerically which describe two-terminal conductance 
and current through the wire. Our exact results may be helpful in 
fabricating mesoscopic or nano-scale spin filter. 
\end{abstract}


\maketitle

\section{Introduction}

Within the last few decades spin polarized transport phenomena~\cite{spin1,
spin2,spin3} in low-dimensional systems have drawn much attention 
due to its potential application in the field of nanoscience and 
nanotechnology~\cite{nano1,nano2}. Discovery of GMR effect~\cite{gmr} 
in Fe/Cr magnetic multilayer in $1980$'s has led to the development of 
a new branch in Condensed Matter Physics - Spintronics, which deals 
with the key idea of exploiting electron spin in transport phenomena. 
The central idea of spintronic applications involves three basic 
steps~\cite{spinsteps1,spinsteps2}, which are injection of spin through 
interfaces, transmission of spin through matter, and finally detection 
of spin. Having considerably larger spin coherence time, quantum 
confined nanostructures such as quantum dots and molecules are therefore 
ideal candidates to study spin dependent transmission which plays a 
significant role for further development in magnetic data storage and 
device processing applications and quantum computation techniques. With 
the increasing interest in generating pure spin current for technological 
purposes, modeling of spin filter is of high importance today.

Till date many theoretical~\cite{theoretical1,theoretical2,theoretical3,
theoretical4,theoretical5,theoretical6,theoretical7,theoretical8} and 
experimental efforts~\cite{experiment1,experiment2} are made to design 
spin filter and increase the efficiency of spin polarization significantly. 
In $2004$, Rokhinson {\em et al.} prepared a spin filter device using GaAs, 
by atomic-force microscopy with local anodic oxidation and molecular beam 
epitaxy methods. They were able to separate charge carriers depending on 
their spin state using the idea of spin-orbit interaction with a weak 
magnetic field. Generation of pure spin current in mesoscopic systems
is a major challenge to us for further advancement in quantum computation.
A more or less common trend~\cite{trend1,trend2} to develop a spin filter 
theoretically is by using ferromagnetic leads or by external magnetic fields. 
But experimental realization of these proposals is somewhat difficult. For
the first case, spin injection from ferromagnetic leads becomes difficult 
due to large resistivity mismatch and for the second one the difficulty
is to confine a very strong magnetic field to a small region like a 
quantum dot (QD). Therefore, attention is being paid for modeling of
spin filter device using the intrinsic properties of quantum 
dots~\cite{intrinsic1,intrinsic2,intrinsic3,intrinsic4,intrinsic5,
intrinsic6}, such as spin-orbit interaction or voltage bias. Ring shaped 
or Aharonov-Bohm (AB) type geometries can achieve high degree of spin 
polarization using Rashba spin-orbit interaction, which lifts the spin 
degeneracy. This has also been achieved by using an AB ring having 
periodic magnetic modulation. 

Aim of the present paper is to study spin dependent transmission through
a magnetic quantum wire which is an array of atomic sites. This system, 
composed of alternately placed magnetic and non-magnetic atoms, 
is attached symmetrically to two non-magnetic (NM) semi-infinite 
one-dimensional ($1$D) electrodes. A simple tight-binding Hamiltonian 
is used to describe the system where all the calculations are done 
by using single particle Green's function formalism~\cite{datta,green1,
green2,green3,green4,green5,green6}. With the help of Landauer 
formula spin dependent conductance is obtained, and the current-voltage 
characteristics are computed from the Landauer-B\"{u}ttiker 
formalism~\cite{land1,land2,land3}. We explore various features of 
spin transport using this simple geometry. Quite interestingly, we 
see that for a certain energy range, transmission probability of up 
spin electron drops to zero, whereas for down spin electrons it 
becomes non-zero and vice-versa. Therefore, tuning the Fermi energy 
$(E_F)$ of the system, the magnetic quantum wire can be used as a 
spin filter for a wide range of applied bias voltage depending on the
strength of localized magnetic moments in the wire. 

The scheme of the paper is as follow. With a brief introduction 
(Section I), in Section II, we describe the model and theoretical 
formulations for the calculation. Section III explores the significant
results which explain the filtering action, and finally, we conclude our 
study in Section IV.

\section{Model and synopsis of the theoretical background}

\begin{figure}[ht]
{\centering \resizebox*{7.8cm}{4.5cm}{\includegraphics{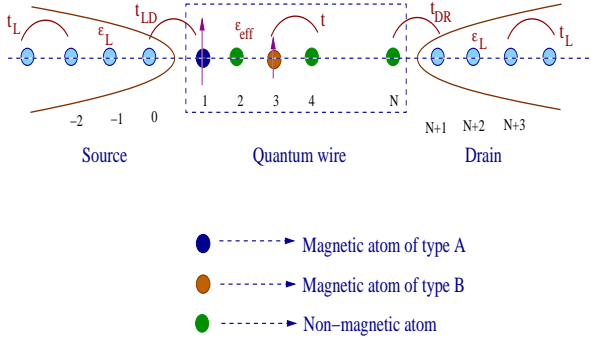}}\par}
\caption{(Color online). A magnetic quantum wire (framed region) of
$N$ atomic sites attached to two semi-infinite $1$D NM electrodes, 
namely, source and drain. Filled green circles correspond to NM atomic 
sites, while the filled blue and brown circles represent atomic sites 
having unequal magnetic moments.}
\label{device}
\end{figure}

The schematic representation of our model is depicted in Fig.~\ref{device}. 
In this figure we illustrate the nano-structure through which spin 
dependent transport is investigated. We study spin transmission through 
a quantum wire of $N$ atomic sites composed of alternately placed 
magnetic and non-magnetic atoms. The wire is attached symmetrically 
to two non-magnetic semi-infinite $1$D metallic electrodes termed 
as source and drain. The atomic sites forming the device are of $3$ 
different types. One of them being non-magnetic and the other two 
being magnetic of types A and B, having two different values of 
localized magnetic moments, $h_A$ and $h_B$, associated with them. 
The orientation of the local moments associated with each 
magnetic site is specified by angles $\theta_n$ and $\phi_n$ ($n$ 
denotes the $n$-th site) in spherical polar coordinate system.
The two metallic electrodes consist of infinite number of non-magnetic 
atoms labeled as $0$, $-1$, $-2$, $\ldots$, $-\infty$ for the left 
electrode and $(N+1)$, $(N+2)$, $(N+3)$, $\ldots$, $\infty$ for the 
right one.

For the whole system (source-wire-drain) we can write the Hamiltonian as,
\begin{equation}
H=H_{W}+H_L+H_R+H_{LW}+H_{WR}
\label{equ1}
\end{equation}
where, $H_{W}$ corresponds to the Hamiltonian of the wire. $H_{L(R)}$ 
represents the Hamiltonian for the left (right) electrode, and 
$H_{LW(WR)}$ is the Hamiltonian representing the wire-electrode coupling.

The spin polarized Hamiltonian for the quantum wire can be written in 
effective one-electron approximation, within the tight-binding formalism 
in Wannier basis, using nearest-neighbor approximation as,
\begin{eqnarray}
H_{W} & = & \sum_{n=1}^N {\bf c_n^{\dagger} \left(\epsilon_0
-\vec{h_n}.\vec{\sigma} \right) c_n}  \nonumber \\
& & + \sum_{i=1}^N 
{\bf \left(c_i^{\dagger}tc_{i+1} + h.c. \right)}
\label{equ2}
\end{eqnarray}
where, \\
${\bf c_n^{\dagger}}=\left(\begin{array}{cc}
c_{n \uparrow}^{\dagger} & c_{n \downarrow}^{\dagger} \end{array}\right)$ \\
${\bf c_n}=\left(\begin{array}{c}
c_{n \uparrow} \\
c_{n \downarrow}\end{array}\right)$ \\
${\bf \epsilon_0}=\left(\begin{array}{cc}
\epsilon_0 & 0 \\
0 & \epsilon_0 \end{array}\right)$ \\
${\bf t}=t\left(\begin{array}{cc}
1 & 0 \\
0 & 1 \end{array}\right)$ \\
${\bf \vec{h_n}.\vec{\sigma}} = h_n\left(\begin{array}{cc}
\cos \theta_n & \sin \theta_n e^{-i \phi_n} \\
\sin \theta_n e^{i \phi_n} & -\cos \theta_n \end{array}\right)$ \\
~\\
\noindent
First term of Eq.~(\ref{equ2}) represents the effective on-site
energies of the atomic sites in the wire. $\epsilon_0$'s are the 
site energies, while the ${\bf \vec{h_n}.\vec{\sigma}}$ refers 
to the interaction of the spin (${\bf \sigma}$) of the injected electron
with the localized on site magnetic moments. This term is responsible for 
spin flipping at the sites. Second term describes the nearest-neighbor 
hopping strength between the sites of the quantum wire. 

Similarly, the Hamiltonian $H_{L(R)}$ can be written as,
\begin{equation}
H_{L(R)}=\sum_i {\bf c_i^{\dagger} \epsilon_{L(R)} c_i} + \sum_i
{\bf \left(c_i^{\dagger} t_{L(R)} c_{i+1} + h.c. \right)}
\label{equ3}
\end{equation}
where, $\epsilon_{L(R)}$'s are the site energies of the electrodes and 
$t_{L(R)}$ is the hopping strength between the nearest-neighbor sites 
of the left (right) electrode. 

\noindent
Here also, \\
~\\
${\bf \epsilon_{L(R)}}=\left(\begin{array}{cc}
\epsilon_{L(R)} & 0 \\
0 & \epsilon_{L(R)} \end{array}\right)$ 
\vskip 0.2cm
\noindent
${\bf t_{L(R)}}=\left(\begin{array}{cc}
t_{L(R)} & 0 \\
0 & t_{L(R)} \end{array}\right)$ \\
~\\
\noindent
The wire-electrode coupling Hamiltonian is described by,
\begin{equation}
H_{LW(WR)}= {\bf \left(c_{0(N)}^{\dagger} t_{LW(WR)} c_{1(N+1)} + 
h.c.\right)}
\label{equ4}
\end{equation}
where, $t_{LW(WR)}$ being the wire-electrode coupling strength.

In order to calculate the spin dependent transmission probabilities 
and the current through the magnetic quantum wire we use single
particle Green's function technique. Within the regime of coherent
transport and for non-interacting systems this formalism is well applied.

The single particle Green's function representing the full system for 
an electron with spin $\sigma$ is defined as~\cite{datta,green1,green2},
\begin{equation}
\bf{G_{\sigma}}=(\bf{E}-\bf{H_{\sigma}})^{-1}
\label{equ5}
\end{equation}
where, 
\begin{equation}
{\bf{E}} = (\epsilon + i \eta) {\bf{I}}
\label{equ6}
\end{equation}
$\epsilon$ being the energy of the electron passing through the system.
$i \eta$ is a small imaginary term added to make the Green's function 
$(G_{\sigma})$ non-hermitian.

Now $\bf{H_{\sigma}}$ and $\bf{G_{\sigma}}$ representing the Hamiltonian 
and the Green's function for the full system can be partitioned 
like~\cite{datta,green1,green2},
\begin{equation}
\bf{H_{\sigma}}=\left(\begin{array}{ccc}
\bf{H_{L\sigma}} & \bf{H_{LW\sigma}} & 0 \\
\bf{H_{LW\sigma}^\dag} & \bf{H_{W\sigma}} & \bf{H_{WR\sigma}}\\
0 & \bf{H_{WR\sigma}^\dag} & \bf{H_{R\sigma}}\\
\end{array} \right) 
\label{equ7}
\end{equation}
\begin{equation}
\bf{G_{\sigma}}=\left(\begin{array}{ccc}
\bf{G_{L\sigma}} & \bf{G_{LW\sigma}} & 0 \\
\bf{G_{LW\sigma}^\dag} & \bf{G_{W\sigma}} & \bf{G_{WR\sigma}}\\
0 & \bf{G_{WR\sigma}^\dag} & \bf{G_{R\sigma}}\\
\end{array} \right) 
\label{equ8}
\end{equation}
where, $\bf{H_{L\sigma}}$, $\bf{H_{R\sigma}}$, and $\bf{H_{W\sigma}}$ 
represent the Hamiltonians (in matrix form) for the left electrode 
(source), quantum wire and right electrode (drain), respectively. 
$\bf{H_{LW\sigma}}$ and $\bf{H_{WR\sigma}}$ are the matrices for the 
Hamiltonians representing the wire-electrode coupling strength. Assuming 
that there is no coupling between the electrodes themselves, the corner 
elements of the matrices are zero. Similar definition goes for the 
Green's function matrix $G_{\sigma}$ as well.

Our first goal is to determine $\bf{G_{W\sigma}}$ (Green's function for 
the wire only) which defines all physical quantities of interest. 
Following Eq.~(\ref{equ5}) and using the block matrix form of 
$\bf{H_{\sigma}}$ and $\bf{G_{\sigma}}$ the form of $\bf{G_{W\sigma}}$ 
can be expressed as~\cite{datta,green1,green2}, 
\begin{equation}
\bf{G_{W\sigma}}=(\bf{E}-\bf{H_{W\sigma}}-\bf{\Sigma_{L\sigma}}-
\bf{\Sigma_{R\sigma}})^{-1}
\label{equ9}
\end{equation}
where, $\bf{\Sigma_{L\sigma}}$ and  $\bf{\Sigma_{R\sigma}}$ represent
the contact self-energies introduced to incorporate the effects of 
semi-infinite electrodes coupled to the system, and, they are expressed 
by the relations~\cite{datta,green1,green2},
\begin{eqnarray}
\bf{\Sigma_{L\sigma}} & = & \bf{H_{LW\sigma}^{\dag} G_{L\sigma} 
H_{LW\sigma}} \\ \nonumber
\bf{\Sigma_{R\sigma}} & = &  \bf{H_{WR\sigma}^{\dag} G_{R\sigma} 
H_{WR\sigma}}
\end{eqnarray}
Thus the form of self-energies are independent of the nano-structure
itself through which transmission is studied and they completely 
describe the influence of electrodes attached to the system.
Now, the transmission probability $(T_{\sigma})$ of an electron
with spin $\sigma$ is related to the Green's function 
as~\cite{datta,green1,green2},
\begin{equation}
T_{\sigma}={\mbox{Tr}}[\bf{\Gamma_{L\sigma} G^r_{W\sigma} \Gamma_{R\sigma} 
G^a_{W\sigma}}]
\end{equation}
where,  $\bf{G^r_{W\sigma}}$ and  $\bf{G^a_{W\sigma}}$ are the retarded
and advanced single particle Green's functions (for the device only) for 
an electron with spin $\sigma$. $\bf{\Gamma_{L\sigma}}$ and  $\bf{{\Gamma_
{R\sigma}}}$ are the coupling matrices, representing the coupling of the 
magnetic quantum wire to the left and right electrodes, respectively, and 
they are defined by the relation~\cite{datta,green1,green2},
\begin{equation}
\bf{\Gamma_{L\sigma(R\sigma)}} = i[\Sigma^r_{L\sigma(R\sigma)} - 
\Sigma^{a}_{L\sigma(R\sigma)}]
\end{equation}
Here, $\bf{\Sigma^r_{L\sigma(R\sigma)}}$ and $\bf{\Sigma^a_{L\sigma
(R\sigma)}}$ are the retarded and advanced self-energies, respectively, 
and they are conjugate to each other. It is shown in literature by 
Datta {\em et al.} that the self-energy can be expressed as a linear 
combination of a real and imaginary parts in the form,
\begin{equation}
{\bf{\Sigma^r_{L\sigma(R\sigma)}}} = {\bf\Lambda_{L\sigma(R\sigma)}}
- i {\bf\Delta_{L\sigma(R\sigma)}}
\end{equation}
The real part of self-energy describes the shift of the energy levels
and the imaginary part corresponds to broadening of the levels. The 
finite imaginary part appears due to incorporation of the semi-infinite 
electrodes having continuous energy spectrum. Therefore, the coupling 
matrices can be easily obtained from the self-energy expression and is
expressed as,
\begin{equation}
{\bf{\Gamma_{L\sigma(R\sigma)}}}=-2{\bf{Im}} 
{\bf{(\Sigma_{L\sigma(R\sigma)})}}
\label{equ14}
\end{equation}
Considering linear transport regime, conductance $(g_\sigma)$ is obtained
using Landauer formula~\cite{datta,green1,green2},
\begin{equation}
g_{\sigma}=\frac{e^2}{h}T_{\sigma}
\end{equation}
Knowing the transmission probability ($T$) of an electron with spin $\sigma$,
the current ($I_\sigma$) through the system is obtained using 
Landauer-B\"{u}ttiker formalism. It is written in the form~\cite{datta,
green1,green2},
\begin{equation}
I_{\sigma} (V)= \frac{e}{h} \int 
\limits_{-\infty}^{+\infty} 
\left[f_L(E)-f_R(E)\right] T_{\sigma}(E)~dE
\label{equ}
\end{equation}
where, $f_{L(R)}=f(E-\mu_{L(R)})$ gives the Fermi distribution function 
of the two electrodes having chemical potentials $\mu_{L(R)}=E_{F} 
\pm eV/2$. $E_F$ is the equilibrium Fermi energy.

\section{Numerical results and discussion}

All the essential features of spin transport through a magnetic quantum 
wire are studied by performing several numerical calculations. Here, we 
assume that the non-magnetic $1$D metallic electrodes are made 
from identical materials, therefore, the site energies and the 
nearest-neighbor hopping strengths are chosen to be identical for the 
two electrodes. Let us begin our discussion by mentioning the values 
of different parameters used for numerical calculations. We choose 
the quantum wire to be made up of $64$ ($N=64$) sites. The on-site 
energies $(\epsilon_0)$ in the quantum wire are chosen to be $0$. 
Magnitudes of the two different local magnetic moments $h_A$ and $h_B$ 
associated with two types of magnetic sites A and B are set as $1.9$ 
and $0.1$. The hopping strength between 
the nearest-neighbor sites of the magnetic quantum wire and for the 
two NM electrodes are set at $t=3$ and $t_L=t_R=4$, respectively. The 
site energies $(\epsilon_{L(R)})$ of all the sites in the two electrodes 
are fixed at $0$. For the sake of simplicity, we choose the unit where 
$h=c=e=1$. Throughout the analysis we study the basic features of spin 
transport for two distinct regimes of electrode-to-magnetic quantum wire
coupling.

\vskip 0.2cm
\noindent
\underline{Case 1:} Weak-coupling limit.
\vskip 0.1cm
\noindent
This regime is defined by the criterion $t_{LW(WR)}<<t$. In this case,
we choose the values as $t_{LW}=t_{WR}=0.5$. 

\vskip 0.2cm
\noindent
\underline{Case 2:} Strong-coupling limit.
\vskip 0.1cm
\noindent
This limit is described by the condition $t_{LW(WR)} \sim t$. In this 
regime we choose the values of hopping strengths as $t_{LW}=t_{WR}=2.5$.

\subsection{Conductance-energy spectrum}

\begin{figure}[ht]
{\centering \resizebox*{7.75cm}{5cm}{\includegraphics{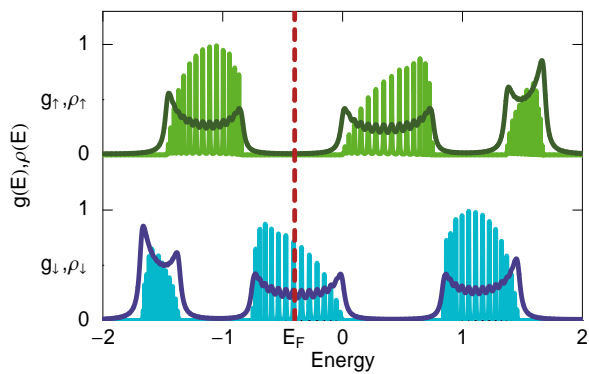}}\par}
\caption{(Color online). $g$-$E$ and $\rho$-$E$ (thick solid lines) curves 
in the limit of weak-coupling for a magnetic quantum wire with $N=64$.
Upper and lower panels correspond to the results of up and down spin
electrons, respectively. Dotted line represents the location of Fermi 
energy $E_F$ of the wire.}
\label{condlow}
\end{figure}

To explain all the relevant features of spin transport we start with 
the conductance-energy characteristics. As illustrative examples, first 
in Fig.~\ref{condlow} we plot the variation of conductance $g$ with 
respect to the energy $E$ for up $(\uparrow)$ and
down $(\downarrow)$ spin electrons separately in the limit of weak 
wire-electrode coupling strength. Also, the variation of average density 
of states (ADOS) are superimposed in each case. The mathematical 
description for the ADOS (symbolized as $\rho(E)$) of the magnetic 
quantum wire including the effect of the two electrodes for an electron 
with spin $\sigma$ is expressed as, 
\begin{equation}
\rho_{\sigma}(E)=-\frac{1}{N \pi} {\mbox{Im}} \left[{\mbox{Tr}}
[{\bf{G_{W\sigma}}}]\right]
\end{equation}
It is observed from the conductance spectra that up and down spin 
electrons follow entirely two different channels while passing through
the wire. This splitting of up and down conduction channels is responsible
for {\em spin filtering} action. Quite interestingly we see that for a 
certain range of energy for which the transmission probability 
$(T_{\uparrow})$ and hence the conductance $(g_{\uparrow})$ for up spin 
electron drops to zero value, shows non-zero transmission probability 
$(T_{\downarrow})$ as well as conductance $(g_{\downarrow})$ due to 
down spin electron. The presence of sharp resonant peaks in the 
conductance spectrum are associated with the energy eigenvalues of 
the full system. In this case due to large system size, $N=64$, the 
sharp resonant peaks get closely spaced to form a quasi band as shown
by the green and blue regions.

\begin{figure}[ht]
{\centering \resizebox*{7.75cm}{5cm}{\includegraphics{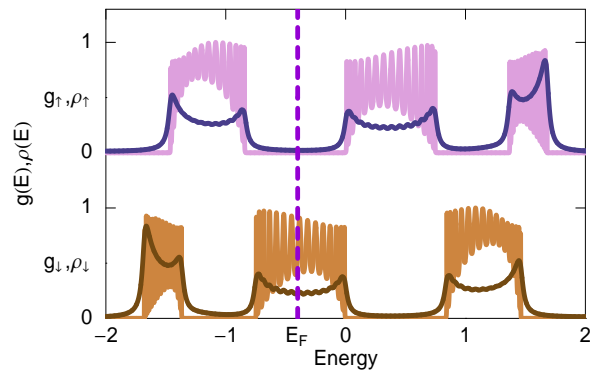}}\par}
\caption{(Color online). $g$-$E$ and $\rho$-$E$ (thick solid lines) curves 
in the limit of strong-coupling for a magnetic quantum wire with $N=64$.
Upper and lower panels correspond to the results of up and down spin
electrons, respectively. Dotted line represents the location of Fermi 
energy $E_F$ of the conductor.}
\label{condhigh}
\end{figure}

In the limit of strong wire-electrode coupling as shown in Fig.~\ref{condhigh}, 
transmission probability $(T_{\sigma})$ almost reaches to the value unity 
and the sharp conductance peaks acquire some broadening which is quantified 
by the imaginary part of the self-energy expression, incorporated to include 
the effect of semi-infinite electrodes. The feature of broadening
is not very clear from this figure due to large system size. Apart from 
increase in transmission probability and broadening of conductance peaks, 
all the other features, as observed in the weak-coupling case, remain 
the same e.g., formation of quasi bands and position of band gaps. 
Therefore, increase in electrode-wire coupling strength does not change 
the position of global gaps in the conductance spectrum, but the coupling
has a strong influence in the study of current-voltage characteristics
as discussed clearly in the references~\cite{green3,green4,green5,green6}.

\subsection{Degree of polarization vs. energy spectrum}

\begin{figure}[ht]
{\centering \resizebox*{7.75cm}{10cm}{\includegraphics{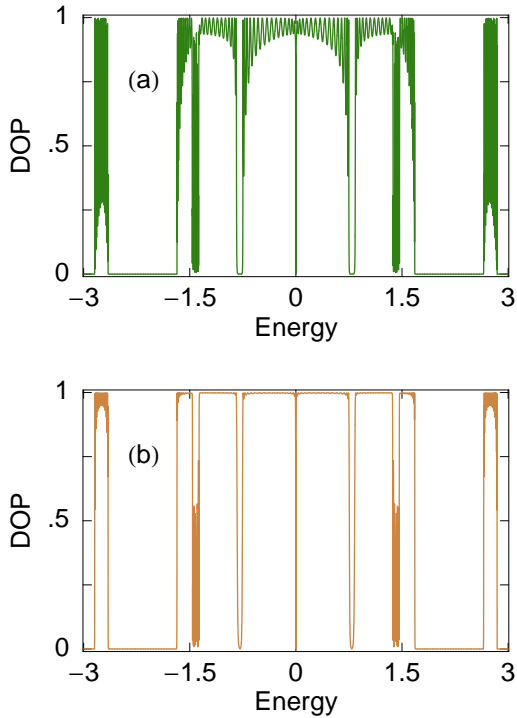}}\par}
\caption{(Color online). Degree of polarization as a function of energy
for a magnetic quantum wire with $N=64$. (a) weak-coupling limit and
(b) strong-coupling limit.}
\label{dop}
\end{figure}

Next, in Fig.~\ref{dop} we show the variation of degree of polarization
(DOP) with respect to the energy of the injected electrons. For a 
particular energy value $E$, the DOP is defined in terms of transmission
probabilities $T_{\sigma}$ in the following way,
\begin{equation}
{\mbox{DOP}} (E)=\left|\frac{T_{\uparrow}(E)-T_{\downarrow}(E)}
{T_{\uparrow}(E)+
T_{\downarrow}(E)}\right|
\end{equation}
DOP gives a quantitative measurement of the spin polarization achieved.
It is observed from Fig.~\ref{dop} that the degree of polarization is 
significantly enhanced with the increase in wire-to-electrode coupling 
strength and for wide range of energies it (DOP) almost reaches to 
the value unity or $100\%$ as expressed conventionally in most of 
the literatures. Thus our proposed model quantum system is a very
good example for designing a {\em spin filter}.

\subsection{Current-voltage characteristics}

\begin{figure}[ht]
{\centering \resizebox*{7.5cm}{4.75cm}{\includegraphics{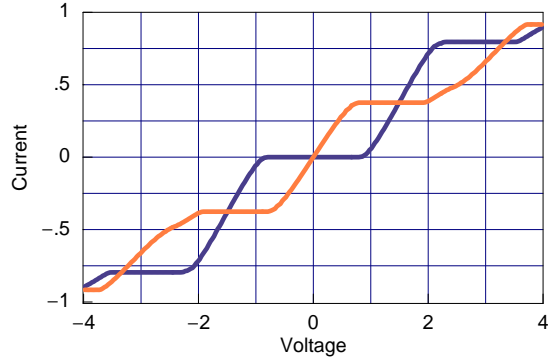}}\par}
\caption{(Color online). Current $I$ as a function of bias voltage $V$ 
in the limit of strong-coupling for a magnetic quantum wire with $N=64$. 
The blue and orange curves correspond to the currents for up and down spin 
electrons, respectively.}
\label{current}
\end{figure}

The spin filtering action becomes clearer in the current-voltage ($I$-$V$)
characteristics presented in Fig.~\ref{current}. Current across the 
magnetic quantum wire is obtained by integrating over the transmission 
function following Landauer-B\"{u}ttiker formalism for a constant Fermi 
energy $(E_F)$. In this case we set the value of $E_F$ at $-0.4$ (dotted
line in Fig.~\ref{condhigh}). As this value of $E_F$ falls in the energy 
gap region of the up spin conductance spectrum, therefore, non-zero 
value of up spin current is obtained after overcoming a finite value
of the applied bias voltage, the so-called threshold voltage $(V_{th})$. 
On the other hand, for any given bias 
voltage $V$, non-zero value of down spin current is observed. Thus spin 
filtering takes place up to the bias voltage $(V_{th})$, when up spin 
current is totally blocked and only down spin current is obtained. In an 
exactly similar way, if we set the Fermi energy at some value of down spin 
energy gap region, down spin current can be blocked totally by passing the 
up spin current only. Here, we plot the current-voltage characteristics 
only in the strong-coupling limit. Exactly a similar filtering action is 
also observed in the case of weak-coupling limit. But the point is that
in the limit of strong-coupling, current amplitude gets magnified 
significantly compared to the weak-coupling limit. The plateau like 
structures in the current-voltage characteristics are observed due to
the presence of global gaps in the conductance spectrum. The value of 
$V_{th}$ depends on the difference between the localized magnetic moments
associated with the magnetic atoms. Therefore, changing the atoms the 
magnitude of $V_{th}$ can be tuned.

\section{Concluding remarks}

To conclude, in the present paper we have investigated spin transport 
through a magnetic quantum wire using single particle Green's function 
formalism. We have adopted a simple tight-binding framework to illustrate 
the system, which is a quantum wire formed by magnetic and non-magnetic 
atomic sites and connected symmetrically to source and drain. We have shown 
the variation of conductance as a function of injecting electron energy
for up and down spin electrons separately for two different strengths
of wire-to-electrode coupling. Conductance spectrum clearly depicts 
the splitting of up and down spin conduction channels which is the key 
idea behind the modeling of a spin filter. Larger the difference between 
the local magnetic moments of the two types of magnetic atoms, smaller 
is the overlap between the up and down conduction channels. Also we have 
plotted the variation of degree of polarization with 
respect to energy for both the coupling regimes. It shows that enhancement 
of coupling strength increases the degree of polarization i.e., improves 
the quality of filtration significantly. Finally, we have obtained the 
current passing through the device using Landauer-B\"{u}ttiker 
formulation. The feature of spin filtering is visualized more prominently 
in the current-voltage characteristics. Tuning the Fermi energy to a 
particular value the device can act as a spin filter i.e., up to a 
certain range of bias voltage only up or down spin current is obtained.

In this work we have calculated all these results by ignoring the effects
of temperature, spin-orbit interaction, electron-electron correlation, 
electron-phonon interaction, disorder, etc. Here, we fix the temperature at 
$0$K, but the basic features will not change significantly even in non-zero 
finite (low) temperature region as long as thermal energy ($k_BT$) is less 
than the average energy spacing of the energy levels of the magnetic
quantum wire. In this model it is also assumed that the two side-attached 
non-magnetic electrodes have negligible resistance.

All these predicted results using such a simple geometry may be useful in
designing a spin polarized source.

\end{document}